\documentclass[submission,PhysCore]{SciPost}

\usepackage{relsize}
\usepackage{physics}
\usepackage{bbold}

\usepackage[utf8]{inputenc} 
\usepackage[T1]{fontenc} 	
\usepackage[english]{babel} 

\usepackage[bitstream-charter]{mathdesign}
\usepackage{geometry} 		
\usepackage{amsmath} 		
\usepackage{mathtools} 		
\usepackage{float} 			
\usepackage{graphicx} 		
\usepackage{tabularx} 		
\usepackage{booktabs} 		
\usepackage{color, xcolor} 	
\usepackage{pdfpages} 		
\usepackage{extarrows} 		
\usepackage{multirow} 		
\usepackage{multicol} 		
\usepackage{enumitem} 		
\usepackage{xspace} 		
\usepackage{stackrel} 		
\usepackage{tikz} 			
\usepackage{braket} 		
\usepackage{bm} 			
\usepackage{tensor} 		
\usepackage{slashed} 		
\usepackage{siunitx} 		
\usepackage{lastpage} 		
\usepackage{cite} 			
\usepackage[normalem]{ulem} 
\usepackage{fontawesome} 	
\usepackage{tocloft} 		
\usepackage{titlesec} 		
\usepackage{doi} 			
\usepackage{hyperref} 		
\usepackage[most]{tcolorbox} 					
\usepackage[nameinlink, capitalize]{cleveref} 	
\usepackage[nottoc, notlot, notlof]{tocbibind} 	
\usepackage[ruled, vlined]{algorithm2e} 		
\usepackage{makecell}
\usepackage[makeroom]{cancel}
\usepackage{feynmf}
\usepackage{aas_macros}          

\makeatletter
\def\BState{\State\hskip-\ALG@thistlm}
\makeatother

\makeatletter
\@ifundefined{pdfoutput}{}{\DeclareGraphicsRule{*}{mps}{*}{}}
\makeatother

\makeatletter
\DeclareRobustCommand*{\bfseries}{%
   \not@math@alphabet\bfseries\mathbf
   \fontseries\bfdefault\selectfont
   \boldmath
}
\makeatother

\hypersetup{
	pdftitle={},
	pdfauthor={Schosser et al.},
	colorlinks=true, 			
	linkcolor={red!50!black}, 	
	citecolor={blue!50!black}, 	
	urlcolor={blue!80!black} 	
} 

\DeclareSymbolFont{usualmathcal}{OMS}{cmsy}{m}{n}
\DeclareSymbolFontAlphabet{\mathcal}{usualmathcal}

\SetArgSty{textnormal}
\SetKwComment{Comment}{{\small\#}~}{}
\SetCommentSty{mycommfont}

\setitemize{itemsep=0pt, parsep=0pt} 				
\setenumerate{itemsep=0pt, parsep=0pt} 				
\setitemize{itemsep=2pt,topsep=2pt,parsep=0pt,partopsep=0pt,leftmargin=*}
\setenumerate{itemsep=0pt,topsep=2pt,parsep=0pt,partopsep=0pt,labelindent=3pt,leftmargin=*}
\usepackage{amsmath}
 
\usepackage{amsthm} 		
\theoremstyle{definition}

\DeclareSymbolFont{usualmathcal}{OMS}{cmsy}{m}{n}
\DeclareSymbolFontAlphabet{\mathcal}{usualmathcal}

\definecolor{red_cb}{HTML}{e41a1c}
\definecolor{blue_cb}{HTML}{377eb8}
\definecolor{green_cb}{HTML}{4daf4a}
\definecolor{purple_cb}{HTML}{984ea3}
\definecolor{orange_cb}{HTML}{ff7f00}

\definecolor{EmeraldGreen}{HTML}{1ea78d}
\definecolor{EnglishRed}{HTML}{b02427}
\hypersetup{colorlinks=true,urlcolor=EmeraldGreen,citecolor=EmeraldGreen,linkcolor=EnglishRed}

\newcommand{\ie}{\text{i.e.}\;}

\renewcommand{\mod}{\theta}

\newcommand{\kl}{D_\text{KL}}

\newcommand{\XLangle}{\Bigl\langle}
\newcommand{\XRangle}{\Bigr\rangle}

\newcommand{\XXXLangle}{\Biggl\langle}
\newcommand{\XXXRangle}{\Biggr\rangle}

\newcommand{\qqquad}{\qquad\quad}

\newcommand\one{\leavevmode\hbox{\small1\normalsize\kern-.33em1}}

\newcommand{\loss}{\mathcal{L}} 	

\newcommand{\arXiv}[2][]{%
	\ifthenelse{\equal{#1}{}}%
	{\href{http://arxiv.org/abs/#2}{arXiv:#2}}%
	{\href{http://arxiv.org/abs/#2}{arXiv:#2~[#1]}}}

\def\slashchar#1{\setbox0=\hbox{$#1$}           
   \dimen0=\wd0                                 
   \setbox1=\hbox{/} \dimen1=\wd1               
   \ifdim\dimen0>\dimen1                        
      \rlap{\hbox to \dimen0{\hfil/\hfil}}      
      #1                                        
   \else                                        
      \rlap{\hbox to \dimen1{\hfil$#1$\hfil}}   
      /                                         
   \fi}

\newcommand{\tikznode}[2]{%
\ifmmode%
\tikz[remember picture,baseline=(#1.base),inner sep=0pt] \node (#1) {$#2$};%
\else
\tikz[remember picture,baseline=(#1.base),inner sep=0pt] \node (#1) {#2};%
\fi}

\def\mathswitchr#1{\relax\ifmmode{\mathrm{#1}}\else$\mathrm{#1}$\xspace\fi}
\def\mathswitch#1{\relax\ifmmode#1\else$#1$\xspace\fi}

\graphicspath{{./figs/}}
\begin{document}


\vspace*{-2.5em}
\hfill{}
\vspace*{0.5em}

\begin{center}{\Large \textbf{
Optimal, fast, and robust inference of \\reionization-era cosmology with the 21cmPIE-INN 
}}\end{center}

\begin{center}
  Benedikt Schosser\textsuperscript{1,2},
  Caroline Heneka\textsuperscript{1}, and
  Tilman Plehn\textsuperscript{1,3}
\end{center}

\begin{center}
{\bf 1} Institut für Theoretische Physik, Universität Heidelberg, Germany\\
{\bf 2}  Astronomisches Rechen-Institut, Zentrum für Astronomie der Universität Heidelberg, Germany\\
{\bf 3} Interdisciplinary Center for Scientific Computing (IWR), Universit\"at Heidelberg,
Germany
\end{center}

\begin{center}
\today
\end{center}


\section*{Abstract}
{\bf
Modern machine learning will allow for 
simulation-based inference from reionization-era 21cm observations at the Square Kilometre Array. 
Our framework combines a convolutional summary network 
and a conditional invertible network through a physics-inspired latent 
representation. 
It allows for an efficient and extremely fast determination of the 
posteriors of astrophysical and cosmological parameters, jointly with well-calibrated and on average unbiased summaries.
The sensitivity to non-Gaussian information makes 
our method a promising alternative to the established power spectra.}

\vspace{10pt}
\noindent\rule{\textwidth}{1pt}
\tableofcontents\thispagestyle{fancy}
\noindent\rule{\textwidth}{1pt}
\vspace{10pt}

\clearpage
\section{Introduction}
\label{sec:intro}

Cosmic Dawn (CD) and the Epoch of Reionization (EoR) mark the emergence of the
first galaxies and stars and the ionization of the intergalactic
medium (IGM) by early luminous sources. Investigating these epochs,
spanning redshifts from $\sim$5--6 to 20, helps us 
understand galaxy evolution, cosmological structure formation,
the thermal history of the Universe, possible primordial sources of
radiation, and the interplay between radiation, gas dynamics, and dark
matter in the formation and evolution of observed structures.

An exciting way to explore the CD and EoR is through the
redshifted 21cm line from the forbidden spin-flip
transition of neutral hydrogen (HI). It is unique
in its sensitivity to the
spatial distribution of neutral hydrogen and the ionization state of
the IGM and offers an exceptional avenue for mapping the large-scale
structure. Experiments such as the Low Frequency Array
(LOFAR)~\cite{van_Haarlem_2013}, the Murchison Widefield Array
(MWA)~\cite{Tingay_2013}, the Hydrogen Epoch of Reionization Array
(HERA)~\cite{DeBoer_2017}, and the Precision Array for Probing the
Epoch of Reionization (PAPER)~\cite{Parsons_2010} strive for a
statistical detection of the 21cm signal, while the Square Kilometre
Array (SKA)\footnote{https://www.skatelescope.org/} promises
3D-tomography.

Intensity mapping of the 21cm line by SKA will allow investigations 
of dark energy and modifications of
gravity~\cite{Heneka:2018kgn,LiuWP2019,Liu:2020JCAP,Berti:2022},
inflation~\cite{Pourtsidou:2016,Modak:2210.05698}, and dark
matter~\cite{Evoli:2014,List2020,Jones:FDM2021}. Here, 
innovative analysis methods are essential, given the inherent non-Gaussianity, foreground
contamination, and systematics. To simulate the 
21cm signal during CD and EoR fast simulation frameworks 
are available for different astrophysical~\cite{Mes07,Mesinger2010,Hassan:2016,Hutter:2021} 
and cosmological scenarios~\cite{Heneka2018}. They are 
complemented by, albeit smaller, databases of radiative hydrodynamical simulations~\cite{Meriot:2023}.

Recent progress in machine learning is transforming data-intensive
analyses in fundamental physics and cosmology~\cite{Dvorkin:2022,Plehn:2022ftl,Moriwaki:2023}. This is especially true
when we can use simulations to relate fundamental
parameters to observations and employ simulation-based
inference~\cite{Brehmer:2019xox,Dax:2021tsq,Huertas-Company:2022wni,Butter:2022rso,Villaescusa-Navarro:2022}.
Traditional simulation-based inference relies on pre-defined high-level 
observables, evaluated as one-dimensional or at most
low-dimensional histograms. This bottleneck prevents us from using the full power of 
measurements or observations.  
The expected size of the SKA dataset, hundreds of Petabytes per year archived, 
and its complexity makes SKA a perfect
example for the need to analyze data without this bottleneck. 

For large cosmological surveys such as SKA we already know that 
convolutional neural networks (CNNs) outperform standard methods for source
detection and characterization~\cite{Boucaud:2020,Hartley:2023}, classification~\cite{Lukic:2018,Mohan:2022,Zhong:2023}, 
and are able to jointly derive astrophysical and cosmological properties
without summary
statistic~\cite{Gillet:2019,HassanCNN:2020}. 
First steps towards 
simulation-based 21cm inference include variational
inference~\cite{hortua:2020} and direct density
estimation~\cite{Zhao:2022,prelogovic:2023,Saxena:2023tue}, including alternative approaches
like wavelet transforms~\cite{Zhao:2023,Zhao:2023b}. 
We show how coupling the 3D-21cmPIE-Net
feature extraction~\cite{Neutsch:2022hmv,Heneka:2023} with a conditional 
invertible neural network (cINNs)~\cite{Ardizzone:2018vuo,Ardizzone:2019xgg}
allows for an optimal, fast, and robust inference of astrophysical and 
cosmological parameters. For the first time in this application, we jointly train these networks, enabling them to mutually refine their representations and achieve more reliable and well-calibrated posteriors. 

We start by introducing the 21cm light cone dataset, the neural posterior estimation (NPE) method
for simulation-based inference, the combination with the 3D-21cmPIE-Net
feature extractor, the physics-inspired training protocol, and a sizeable
range of validation and quality control methods in Sec.~\ref{sec:inf}.
In Sec.~\ref{sec:res_cal} we use marginalized 1-dimensional posteriors to
control and confirm the calibration and the robustness of our 
21cmPIE-INN setup. In Sec.~\ref{sec:res_mock} we show that its controlled,
excellent performance remains when we add noise to the pure simulations
for a realistic mock dataset.
Finally, in Sec.~\ref{sec:res_inf} we show how the 21cmPIE-INN can extract 
a multi-dimensional posterior for astrophysics and cosmological parameters 
from a single light cone. Additional quality control measures and posteriors 
for more light cones are given in the Appendix.

\section{Inference for 21cm tomography}
\label{sec:inf}

Our goal is to use simulation-based inference to extract as much
information as possible from complex 21cm data and avoid the
bottlenecks of classic analysis methods. This requires an appropriate
representation of the data, introduced in Sec.~\ref{sec:inf_data}, the
conditional generative neural network described in
Sec.~\ref{sec:inf_bf}, a physics-inspired data pre-processing introduced
in Sec.~\ref{sec:inf_pie}, a dedicated training
protocol discussed in Sec.~\ref{sec:inf_train}, and a detailed 
validation, Sec.~\ref{sec:inf_valid}. Our framework will then allow 
for a fast, amortized inference of cosmological and astrophysical parameters
from a single 21cm light cone.

\subsection{21cm light cone data}
\label{sec:inf_data}

Our data consists of 5000 3D-light cones (LC) of 21cm brightness
temperature fluctuations $\delta T_b(x,\nu)$, with on-sky coordinates
$x$ and frequency $\nu$. The LCs are produced with the semi-numerical
code
21cmFASTv3~\cite{Murray2020}.\footnote{https://github.com/21cmFAST/21cmFAST}
It generates initial density and velocity conditions and evolves them
at first and second order perturbation theory using the Zel'dovich
approximation~\cite{zeldovich}. A region is flagged as ionized, if the
fraction of collapsed matter, $f_\text{coll}$, exceeds the inverse
ionizing efficiency of star formation, $\zeta^{-1}$. The fraction
$f_\text{coll}$ is calculated in an excursion-set approach, where the
density field is filtered with a top-hat of decreasing size. The code
accounts for partially ionized regions with an ionized fraction
$f_\text{coll} \zeta$.

Besides the ionization fraction, the 21cm signal at higher redshifts
crucially depends on the spin gas temperature $T_\text{S}$, which in
turn depends on couplings to kinetic gas temperature and density.  We
do not assume the so-called post-heating regime and instead fully
evolve spin temperature boxes. To generate LCs, coeval cubes of 21cm
brightness temperature fluctuations, evolved with redshift, are
stitched together in the last step.

The resulting 21cm brightness fluctuations depend on several
cosmological and astrophysical parameters. For our simple,
proof-of-concept study we combine two parameters defining our
cosmological model, two parameters describing astrophysics during
cosmic dawn, and two parameters to account for EoR
astrophysics~\cite{Neutsch:2022hmv}:
\begin{itemize}
\item Matter density $\Omega_\text{m} \in [0.2,0.4]$\\ It
  controls structure formation, where a wide range encompasses the
  Planck limits~\cite{Planck:2018vyg};
\item Warm dark matter mass $m_\text{WDM}\in
  [0.3,10]\,\text{keV}$\\ The conservative limit allows for a wide
  range of possible behavior, where the lower limit exhibits a
  tension with Cold Dark Matter (CDM), and current astrophysical constraints point
  towards values larger than a few keV~\cite{Villasenor:2023,Irsic:2023}. Here, structure formation
  looks more and more similar to CDM, as the free-streaming length is
  inversely proportional to the WDM mass;
\item Minimum virial temperature $T_\text{vir} \in
  [10^4,10^{5.3}]\,\text{K}$\\ The minimum virial temperature needed
  for cooling within halos to enable star formation;
\item Ionization efficiency $\zeta\in [10,250]$\\ It is represented
  by the composite parameter
  \begin{align}
  \zeta = 
  30 \frac{f_\text{esc}}{0.3} \; 
  \frac{f_\star}{0.05} \; 
  \frac{N_{\gamma/b}}{4000} \; 
  \frac{2}{1+n_\text{rec}} \; ,
  \end{align}
  where factors such as the escape fraction of ionizing photons into
  the intergalactic medium $f_\text{esc}$, the fraction of galactic
  gas in stars $f_\star$, the number of ionizing photons per baryon
  in stars $N_{\gamma/b}$, and the typical number density of
  recombinations for hydrogen in the intergalactic medium 
  $n_\text{rec}$ contribute to a versatile range of recombination
  scenarios;
\item Specific X-ray luminosity $L_\text{X} \in
  [10^{38},10^{42}]\,\text{erg}\,\text{s}^{-1}\,\text{M}_\odot^{-1}
  \,\text{yr}$\\ Integrated luminosity $<2\,\text{keV}$ per unit star
  formation rate that escapes host galaxies;
\item X-ray energy threshold for self-absorption by host galaxies $E_0
  \in [100,1500]\,\text{eV}$\\ X-rays with energies below $E_0$ do
  not escape the host galaxy.
\end{itemize}
All other cosmological parameters are fixed to the Planck
measurements~\cite{Planck2018}, assuming flatness and a cosmological
constant. This means $\Omega_\text{b}=0.04897$, $\sigma_8 = 0.8102$,
$h=0.6766$, and $n_s=0.9665$.

To generate our training data, we randomly sample parameters
from flat priors and then simulate the corresponding light cone of
21cm brightness offset temperature. The box size is $200\,\text{Mpc}$
at a resolution of $1.42\,\text{Mpc}$, and the redshift range
simulated is $z=5~...~35$. Each light cone has the shape of
$(2350,140,140)$ voxels, keeping in mind that $\Omega_\text{m}$
impacts the length of each light cone in terms of redshift. Hence,
only at $\Omega_\text{m}=0.4$ the simulated light cone includes
$z=35$, light cones for smaller $\Omega_\text{m}$ stop at slightly
lower redshifts, to keep the number of pixels in temporal or redshift
direction fixed.

Due to the wide parameter ranges some of the simulated reionization
histories and light cones are excluded by observations. We filter the
light cones to exclude extreme reionization histories, requiring that
the optical depth $\tau_\text{reio}$ is within $5\sigma$ of the Planck
measurement $0.054\pm 0.007$~\cite{Planck:2018vyg}, and that the IGM
mean neutral fraction at redshift 5 is below 0.1.  From 5000 valid
light cones in our dataset, we use 3600 for training, 400 for
validation, and 1000 for testing the network.

Going beyond the idealized, pure simulations, we generate mock observed light
cones using
21cmSense~\cite{Pober_2013,Pober_2014}.\footnote{https://github.com/rasg-affiliates/21cmSense}
Our 5000 simulated light cones are transformed by coevally
evolved simulation boxes at fixed redshifts. These boxes are split at
certain redshift values, and thermal noise is calculated for each box
using 21cmSense. The resulting noise is added to the
Fourier-transformed box, and the mock light cone is reconstructed in
real space. The thermal noise assumptions are based on 1080 hours of
integrated SKA-Low stage~1 observations with specific instrument
characteristics and baseline distribution. Three foreground settings
(optimistic, moderate, pessimistic) in 21cmSense account for different
scenarios, where the optimistic scenario considers the 21cm foreground
wedge in $k$-space covering only the primary field-of-view of the
instrument. In this work, we use this optimistic foreground
scenario.  Mock light cones with this noise setting are mainly
affected at higher redshifts. 

\subsection{Neural Posterior Estimation}
\label{sec:inf_bf}

Conditional generative networks have shown great promise for Bayesian
inference in fundamental physics and
cosmology~\cite{Bieringer:2020tnw,Kang2023,Eisert2023,adler2023applicationdriven}. NPE allows
for simulation-based inference without assuming a likelihood shape. It
combines a summary network, to reduce simulated or observed data to an
appropriate latent representation, with a conditional generative
network. Specifically, we implemented the invertible neural network
(INN)~\cite{Ardizzone:2018vuo} version of a normalizing
flow~\cite{Winkler:2019lio,Papamakarios:2019fms}, following the
BayesFlow~\cite{radev2020bayesflow,radev2023bayesflow} concept. Our implementation is available to use on \href{https://github.com/cosmostatistics/21cm_pie}{GitHub}.
Conditional on an
observation, this cINN generates the posterior in parameter space. For
21cm tomography, its input is the full signal light cones of the 21cm
offset brightness temperature, without reducing the 3D-maps to
summary statistics, like the power spectrum.

\begin{figure}[t]
    \centering
    \includegraphics[width=0.9\textwidth]{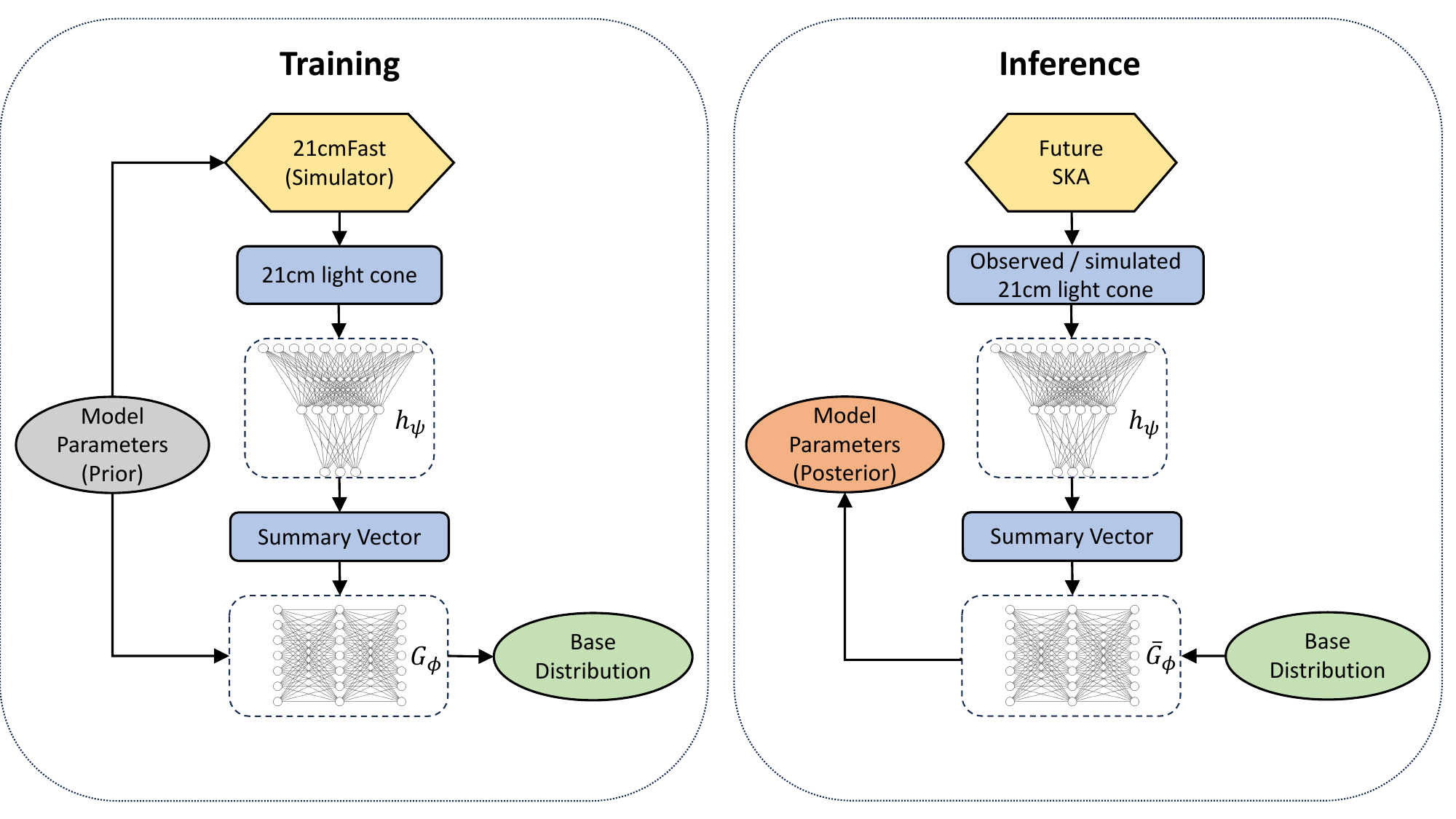}
    \caption{Illustration of Bayesian inference with
      conditional generative networks, following the 
      BayesFlow concept.} 
    \label{fig:bayesflow_concept}
\end{figure}

This inference method is illustrated in
Fig.~\ref{fig:bayesflow_concept}. The two networks are trained on
simulations, providing the model parameters and paired simulated
data. The simulated data is passed through the summary network, to
provide the condition for the cINN. This cINN maps the model
parameters to a Gaussian latent distribution by minimizing a
likelihood loss.  For the inference, the summarized data is
again the condition for the cINN, which now samples from the Gaussian
to generate the posterior in model space.

The simulation uses the model parameters $\mod_{1:D}$ to generate
data $x_{1:N}$, where the elements $x_i$ can be scalars or vectors.
To simplify our notation, we first omit the summary network, so the
data $x$ is fed directly to the cINN. This cINN links a latent
Gaussian distribution and the posterior over model
space~\cite{Plehn:2022ftl},
\begin{align}
\text{latent $r \sim \mathcal{N}_{0,1}$} 
\quad
\stackrel[\leftarrow \; \overline{G}_\phi(\mod|x)]{G_\phi(r|x) \rightarrow}{\xleftrightarrow{\hspace*{1.5cm}}}
\quad
\text{model space $\mod \sim p(\mod|x)$} \; .
\label{eq:cinn_mapping}
\end{align}
Here $\overline{G}_\phi(\mod|x)$ denotes the inverse transformation to
$G_\phi(r|x)$, both encoded in the network parameters $\phi$.  The
training goal is to approximate the true posterior,
\begin{align}
p_\phi(\mod|x)\approx p(\mod|x)
\end{align} 
for all possible parameters $\mod$ and data $x$. To this end, we
minimize the Kullback-Leibler (KL) divergence between the approximate
and true posteriors 
\begin{align}
  \kl[p(\mod|x),p_\phi(\mod|x)] 
  &= \XLangle \log p(\mod|x)- \log p_\phi(\mod|x) \XRangle_{p(\mod,x)} \; ,
  \label{eq:kl}
\end{align}  
or the weight-dependent loss function
\begin{align}
  \loss_\text{cINN} = - \XLangle \log p_\phi(\mod|x) \XRangle_{p(\mod,x)} \; .
  \label{eq:loss1}
\end{align}
This loss function is evaluated over pairs of model parameters and the
corresponding simulated data. The posterior is encoded in the cINN
through a Jacobian of $\overline{G}_\phi$,
\begin{align}
  p_\phi(\mod|x)
  &= \mathcal{N}_{0,1}(\overline{G}_\phi(\mod|x)) \;
  \left| \pdv{\overline{G}_\phi(\mod|x)}{\mod} \right| \notag \\
  \Rightarrow \qquad   
  \loss_\text{cINN}
  &= - \XXXLangle
  \log \mathcal{N}_{0,1}(\overline{G}_\phi(\mod|x))
  + \log \left| \pdv{\overline{G}_\phi(\mod|x)}{\mod} \right|
  \XXXRangle_{p(\mod,x)} \notag \\
  &= - \XXXLangle
  \frac{|\overline{G}_\phi(\mod|x)|^2}{2}
  + \log \left| \pdv{\overline{G}_\phi(\mod|x)}{\mod} \right|
  \XXXRangle_{p(\mod,x)} \; .
  \label{eq:loss2}
\end{align}
The first term regularizes the network, while the second term trains
the Jacobian of the cINN.

The summary network $h_\psi$ transforms the input data before it
enters the cINN as the condition. It does not have to be big and can
be trained together with the cINN, using the NPE loss function
\begin{align}
  \loss_\text{NPE}
  &= - \XXXLangle
  \frac{|\overline{G}_\phi(\mod|h_\psi(x))|^2}{2}
  + \log \left| \pdv{\overline{G}_\phi(\mod|h_\psi(x))}{\mod} \right|
  \XXXRangle_{p(\mod,x)} \; .
  \label{eq:loss3}
\end{align}
The number of instances we train the network on is free, as long as we
evaluate the network on the same number of instances of the observed
or test data $x^o_{1:N}$.

The loss in Eq.\eqref{eq:loss2} assumes that the Jacobian relating the
model parameters $\mod$ and latent random variables $r$ can be
evaluated fast~\cite{9089305_normflow}. The classic choice is a stack
of affine coupling blocks~\cite{dinh2017density} and rotational
layers. These affine layers are simple and extremely fast. In case we
need a more expressive invertible network, we can replace them by
cubic~\cite{durkan2019cubic} or rational quadratic
splines~\cite{durkan2019neural}, if needed with learned
rotations~\cite{Heimel:2022wyj} or periodic boundary
conditions~\cite{Butter:2022vkj}.

\subsection{21cmPIE-INN}
\label{sec:inf_pie}

\begin{figure}[b!]
    \centering
    \includegraphics[width=0.9\textwidth]{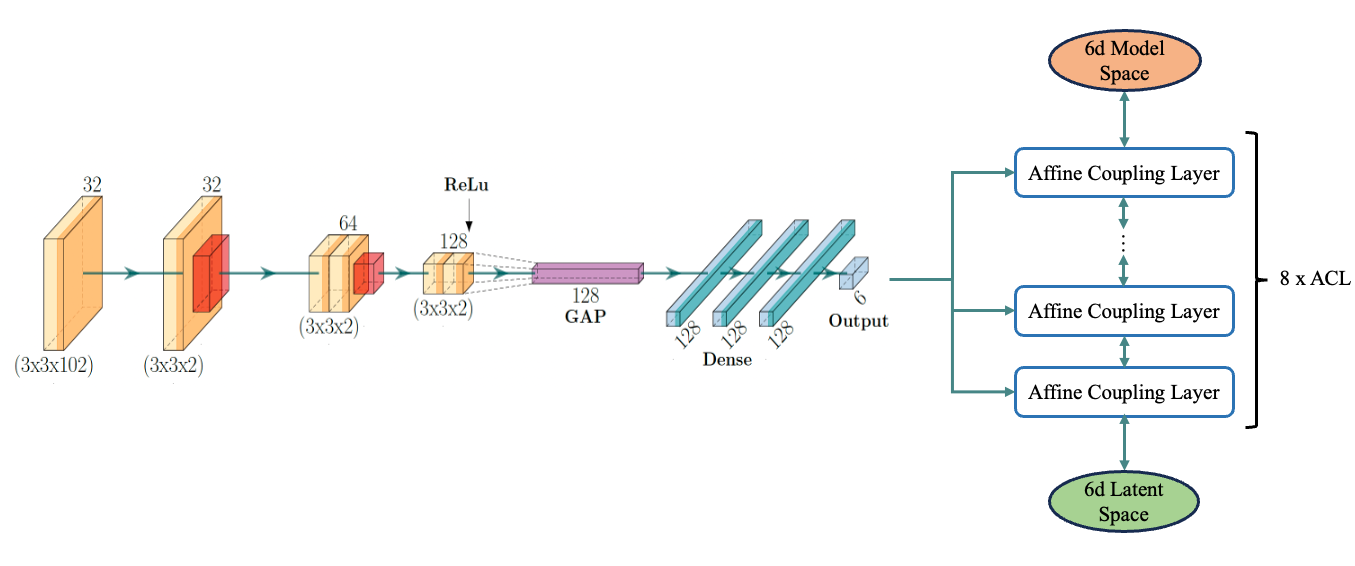}
    \caption{Schematic representation of the 21cmPIE-INN, combining the physics-inspired summary network~\cite{Neutsch:2022hmv} and the cINN.}
    \label{fig:architecture}
\end{figure}

As described above,
our summary network $h_\psi(x)$ compresses a large three-dimensional
data object, the light cone. We choose a strong compression to the six
parameters used for light cone simulation in CD and EoR
astrophysics and cosmology, as described in Sec.~\ref{sec:inf_data}. In
this case, the network needs to be very expressive. We use a 
3D-convolutional network (CNN), the 3D-21cmPIE-Net introduced for
parameter regression~\cite{Neutsch:2022hmv,Heneka:2023}. 
The 3D-21cmPIE-Net provides fast and optimal convergence at a moderate 
size of the required training dataset, compared to alternative networks~\cite{prelogovic:2023,Zhao:2022}. 
It has been shown to
efficiently provide unbiased parameter estimates for both
astrophysical and cosmological parameters, outperforming for example
larger Long Short Term Memory networks, and requiring smaller training datasets.
The architecture
is summarized in Tab.~\ref{tab:architecture} and schematically shown 
in Fig.~\ref{fig:architecture}. The asymmetric $(3\times
3 \times 102)$-kernel of the first filter reflects the difference
between fluctuations in temporal $z$-direction and spatial
direction. The $(1\times 1 \times 102)$-stride reduces the
dimensionality of the following layers, while still capturing the
relevant physics. The kernel size of the hidden layers is set to
$(3\times 3 \times 2)$ and max pooling layers are applied only in the
spatial direction. In front of three fully connected layers is one
global average pooling layer to impede overfitting. The hidden layers
use a ReLU activation function.

\begin{table}[b!]
\centering
\begin{small} \begin{tabular}{l | ll}
\toprule
& Layer & Shape \\
\midrule
\multirow{16}{*}{3D-CNN}
&Input Layer & (1,140,140,2350)\\
&3x3x102 Conv3D & (32,138,138,23)\\
&3x3x2 Conv3D & (32,136,136,22)\\
&2x2x1 Max Pooling & (32,68,68,22)\\
&3x3x2 Conv3D & (64,66,66,21)\\
&1x1x0 Zero Padding & (64,66,66,20)\\
&3x3x2 Conv3D & (64,66,66,20)\\
&2x2x1 Max Pooling & (64,33,33,20)\\
&3x3x2 Conv3D & (128,31,31,19)\\
&1x1x0 Zero Padding & (128,33,33,19)\\
&3x3x2 Conv3D & (128,31,31,18)\\
&Global Average Pooling & (128)\\
& 3 x Dense & (128)\\
&Dense & (6)\\
\midrule
\multirow{3}{*}{cINN}
& Number of inferred parameters & 6\\
&Coupling layers & 8\\
&Fully connected coupling layer architecture & 256, for all layers\\
\bottomrule
\end{tabular} \end{small}
\caption{3D-CNN and cINN architectures and hyperparameters.}
\label{tab:architecture}
\end{table}

The summary network condenses each LC with its complex physics information
to a six-dimensional latent distribution. In the next section, we will see that
the first stage of our training protocol pushes the network to identify this 
vector with the cosmological and astrophysical parameters
\begin{align}
\mod = 
\left\{ \Omega_\text{m}, m_\text{WDM}, T_\text{vir}, \zeta, L_\text{X}, E_0 \right\} \; .
\label{eq:paras}
\end{align}
The cINN uses this physics-inspired latent LC representation as a condition
for linking the six-dimensional model space to a Gaussian of the same
dimensionality. 
Each of the six model parameters in Eq.\eqref{eq:paras}
is normalized to the range $[0,1]$. 

The idea behind this setup is that the summary network extracts the 
relevant physics parameters from the complex data representation, a
standard regression task, and that the cINN only needs to learn the correlations
and the uncertainties for the posterior estimation.
In case this intermediate representation of Eq.\eqref{eq:paras} is not optimal,
the joint training of the summary network and the cINN corrects for a
potential shortcoming to find the optimal latent representation. 
Depending on the downstream task, a higher dimensional latent space could offer additional flexibility. Here, we find the parameter-inspired choice to be sufficient for well-calibrated posteriors as shown in the following.

Nevertheless, this simple cINN
task only requires a small number of simple coupling layers.  We stack
8 affine coupling layers, where each of the internal fully connected
networks has 256 nodes with one hidden layer and ReLU activation. For
the INN we use FrEIA (Framework for Easily Invertible
Architectures)~\cite{freia}, all our training and evaluation is
performed in PyTorch~\cite{pytorch} with the Adam
optimizer~\cite{Kingma:2014vow}.

\subsection{Training}
\label{sec:inf_train}

\begin{figure}[t]
  \centering
  \includegraphics[width=.9\textwidth]{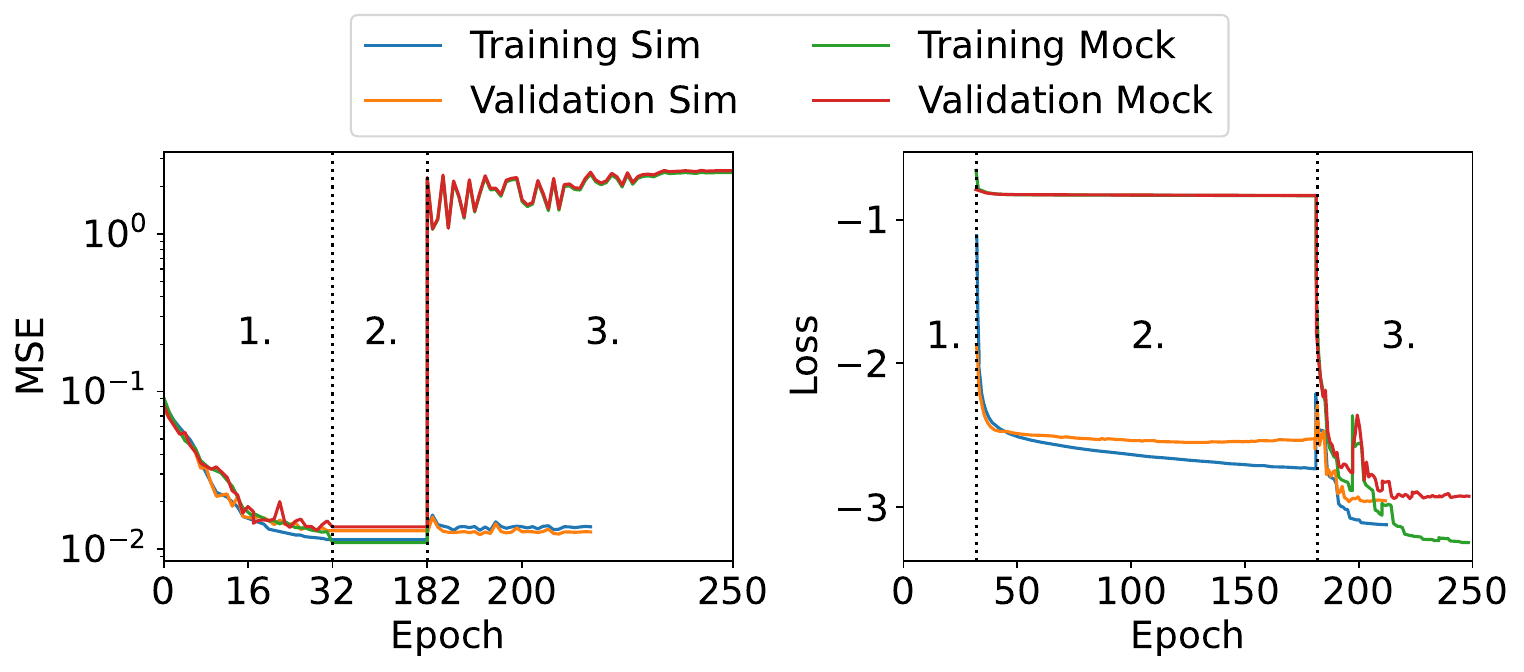}
  \caption{MSE of the summary network (left) and the cINN loss (right) during the three training stages, for training and validation datasets.}
\label{fig:loss}
\end{figure}

For the training, we use 3600 simulated light cones, in 450 batches of
eight. As the numbers of network parameters in the cINN and PIE-Net
differ ($\#$CNN parameters: 651526, $\#$cINN parameters: 33488), the
summary network needs to be pre-trained to provide a sensible basis for
the cINN training. Without the pre-training step, the combination of both networks did not converge when jointly trained, probably due to the large difference in number of network parameters between the two networks. We use the physics information of the astrophysical and cosmological simulation parameters to set up a summary statistics in this region of the latent space. We train two different models, one on the pure noiseless simulations and one on the mock observations. We use a 3-stage training:
\begin{enumerate}
\item First, we pre-train the summary network as a regression network with an MSE-loss, since
  the parameter labels in Eq.\eqref{eq:paras} are known. The learning rate is
  initially set to $4\cdot 10^{-4}$ and halved after 15 and 20 epochs.
  The training ends after 32 epochs. We do not require perfect
  convergence for this first stage.
\item Next, we only train the cINN with the pre-trained summary
  network. The learning rate is constant at $4\cdot 10^{-4}$, and the
  training stops after 150 epochs, when the cINN starts to
  overfit. Adapting the learning right might slightly improve this
  pre-training, but without any effect on the final training.
\item Finally, both networks are trained together, so they can optimize the
  latent representation. The learning rate is reduced by a factor of 0.3 every
  five steps, starting from $4\cdot 10^{-4}$, and
 stays constant after epoch 20. We stop the training 
 after 32 epochs for the simulations and after 68 for the mocks.
\end{enumerate}
Examples of the loss evolution for pure simulation and mock data in the three stages are shown in
Fig.~\ref{fig:loss}. The left panel shows the MSE between the true labels and 
the output of the summary network. During the first stage, it defines the 
training objective and therefore decreases to the point where training and 
validation loss slightly deviate. The second stage does not affect the summary 
network and its latent LC representation. In the third stage, when both networks 
are trained with the cINN likelihood loss in Eq.\eqref{eq:loss3}, the MSE 
remains constant for the pure simulation and increases again for the mock data. 
The reason is that now the 6-dimensional output of the summary network can be corrected,
away from the physical parameters given in Eq.\eqref{eq:paras}, a more informative and unbiased summary 
can be found. For the pure simulation,
such an adjustment is not needed. However, for mock data, the increased MSE indicates
a significant change in the optimal latent representation when the two networks are trained 
jointly in stage 3. This difference is expected from a pure machine learning perspective ---
adding noise to the data makes
it easier for the training to move around the loss landscape, including adjustments 
of the summary network required for optimal inference. We have checked that, indeed, the
inference of cosmological parameters from mock data outperforms the pure simulations.
This is why in the following we consider inference from pure simulations critical for  
understanding the behavior of the networks, but use the mock data to benchmark the 21cmPIE-INN
performance.

The right panel in Fig.~\ref{fig:loss} shows the evolution of the cINN loss from
Eq.\eqref{eq:loss3}. It is not defined in the first stage. In the second stage, it 
decreases and approaches the respective plateaus quickly. However, the loss values
on the plateaus for the pure simulation and the mock data are different. Only when 
we also adjust the summary network in the third stage, the loss for the mock data
reduces to the same level as for the pure simulation, indicating that at this 
stage both setups work at a similar level. As mentioned before, we still stick 
to the more realistic mock data performance whenever possible.

On an NVIDIA GeForce GTX 1080 Ti the training time per epoch is about
1 hour in stage 1, 1 minute in stage 2, and 1.3 hours in stage
3. These times reflect that in stages~1 and~3 the summary network is
updated, so we need to read the LC dataset. As it is large (900~GB), we
we cannot load it into memory and instead read the files for every
batch.  In stage~2 we use the fixed output of the summary network.
The total training time is around 74 hours. The architecture and
hyperparameters are summarized in Tab.~\ref{tab:architecture}.  For
the simulated mock light cones we train another set of identical
networks in the same way.

\subsection{Validation}
\label{sec:inf_valid}

To assess the network performance we use a set of standard
metrics. First, we evaluate the parameter recovery. For a perfect
posterior approximation, the mean of the marginalized posterior should
statistically coincide with the true value of the parameter. To
quantify the deviation, we use two metrics that measure if a sample
of true parameters $\mod_j$ corresponds to a sample of estimated
parameters $\hat{\mod}_j$.  First, the coefficient of
determination
\begin{align}
  R^2
  =1-\sum_{j=1}^J \frac{( \mod_j-\hat{\mod}_j )^2}
  {( \mod_j -\bar{\mod}_j )^2}\, 
\label{eq:R2_determination}
\end{align}
measures the proportion of variance. The estimated parameters are the
means of the sampled parameters and $\bar{\mod}$ denotes the mean of
the true parameter samples.  For perfect parameter recovery, we find
$R^2=1$.  Alternatively, the normalized root mean square deviation
(NRMSE),
\begin{align}
  \text{NRMSE}
  =\frac{\sqrt{\dfrac{1}{J} \sum_{j=1}^J (\mod_j -\hat{\mod}_j )^2}}
  {\mod_\text{max}-\mod_\text{min}}\,,
\label{eq:NRMSE}
\end{align} 
includes a different normalization to make it scale-independent and
to allow for comparison throughout all parameter ranges.

Second, simulation-based calibration (SBC)~\cite{SBC} is a self-consistency
check to visually detect systematic biases. Given a sample from the
prior $\tilde{\mod} \sim p(\mod)$ and one from the forward model
$\tilde{x}\sim p(x|\mod)$ one can integrate out the $\tilde{\mod}$ and
$\tilde{x}$ to recover the prior
\begin{align}
  p(\mod)
  =\int \dd \tilde{x}\dd\tilde{\mod} \; p(\mod|\tilde{x})p(\tilde{x}|\tilde{\mod}) \, .
  \label{eq:sbc}
\end{align}
This equation allows us to check for self-consistent sampling. If one
samples from the correct posterior, Eq.\eqref{eq:sbc} holds for any
form of the posterior. Any violation of this equality indicates a
problem in the sampling. If there is no violation then one can claim that the posterior is on average without bias and uses all the information from the summary. However, different biases in certain regions may cancel each other and the summary might not be optimal. 
The authors of
Ref.~\cite{SBC} propose the SBC algorithm

\begin{center} \begin{small} \begin{tabular}{l l}
         \toprule
         1:&\textbf{for} $m=1,...,M$ \textbf{do}: \\
         2:&\qquad Sample $\tilde{\mod}_m\sim p(\mod)$\\
         3:&\qquad Simulate a dataset $\tilde{x}_m$\\
         4:&\qquad Draw posterior samples $\hat{\mod}_l \sim p_\phi(\mod|\tilde{x}_m)$\\
         5:&\qquad Compute rank statistic $r^{(m)}=\sum_{l=1}^L I_{[\hat{\mod}_l<\tilde{\mod}_l]}$ \\
         6:&\qquad Store $r^{(m)}$ \\
         7:&\textbf{end for}\\
         8:&Create a histogram of $\{r^{(i)}\}_{m=1}^M$ and inspect for uniformity. \\
         \bottomrule
  \end{tabular} \end{small} \end{center}
SBC requires a framework where fast amortized inference is possible, because many, typically up to a thousand or tens of thousands, posteriors need to be averaged. Therefore, this consistency check is only available in SBI and not for MCMC methods as it is computationally not feasible to perform this many MCMC runs for 21cm cosmology. Here, MCMC runs typically take weeks to months, as the signal is modelled based on simulations. SBC is already used in the field of Cosmology, see e.g.~\cite{10.1093/mnras/stad2027}. Intuitively, it counts how many sample values are larger than the true value. Normalized by the total number of samples, this gives the rank of the true parameter within this posterior. Over many iterations this should converge to a uniform distribution and visual discrepancies can be interpreted. If the histogram clusters at the extremes, $\cup$- or $\cap$-shaped, it shows over- or underconfidence, respectively. A slope in the histogram shows a bias.

Third, latent space examination allows us to trace the learned
connection between model parameters and a unit Gaussian during the
training. As training progresses, we can evaluate the cINN in the
forward direction rather than the inference direction, to see if the
latent distribution approaches a Gaussian shape.

Finally, the calibration error~\cite{err_cal} quantifies how well the
coverage of an approximate posterior matches the coverage of an
unknown true posterior. For each parameter, the marginalized
approximate posterior is given and the $\alpha$-credible intervals can
be calculated, with $\alpha \in (0,1)$. For each $\alpha$ the fraction
of true parameter values lying in the interval is denoted with
$\alpha_\mod$. A perfectly calibrated approximate posterior is given
by $\alpha_\mod=\alpha$. The calibration error is then defined as
\begin{align}
  \text{Err}_\text{cal}=\sum_j |\alpha_{\mod,j} - \alpha_j|\, ,
  \label{eq:calibration_error}
\end{align}
with equally spaced $\alpha_j$. A vanishing calibration error
indicates perfect calibration.

\section{Results}
\label{sec:res}

After training the 21cmPIE-INN in three stages, it can be used for
inference, as illustrated in Fig.~\ref{fig:bayesflow_concept}. This NPE setup extracts the full posterior by first reducing
the dimensionality of the LC to a summary vector and then sampling
from a Gaussian into model parameter space, conditioned on this summary
vector. By examining the posterior and comparing it with the true
labels, we can determine the performance of the network.  We always
discuss the results on the pure simulations first, followed by
mock data including noise.

\subsection{Performance and calibration}
\label{sec:res_cal}

\begin{figure}[t]
    \includegraphics[width=\textwidth]{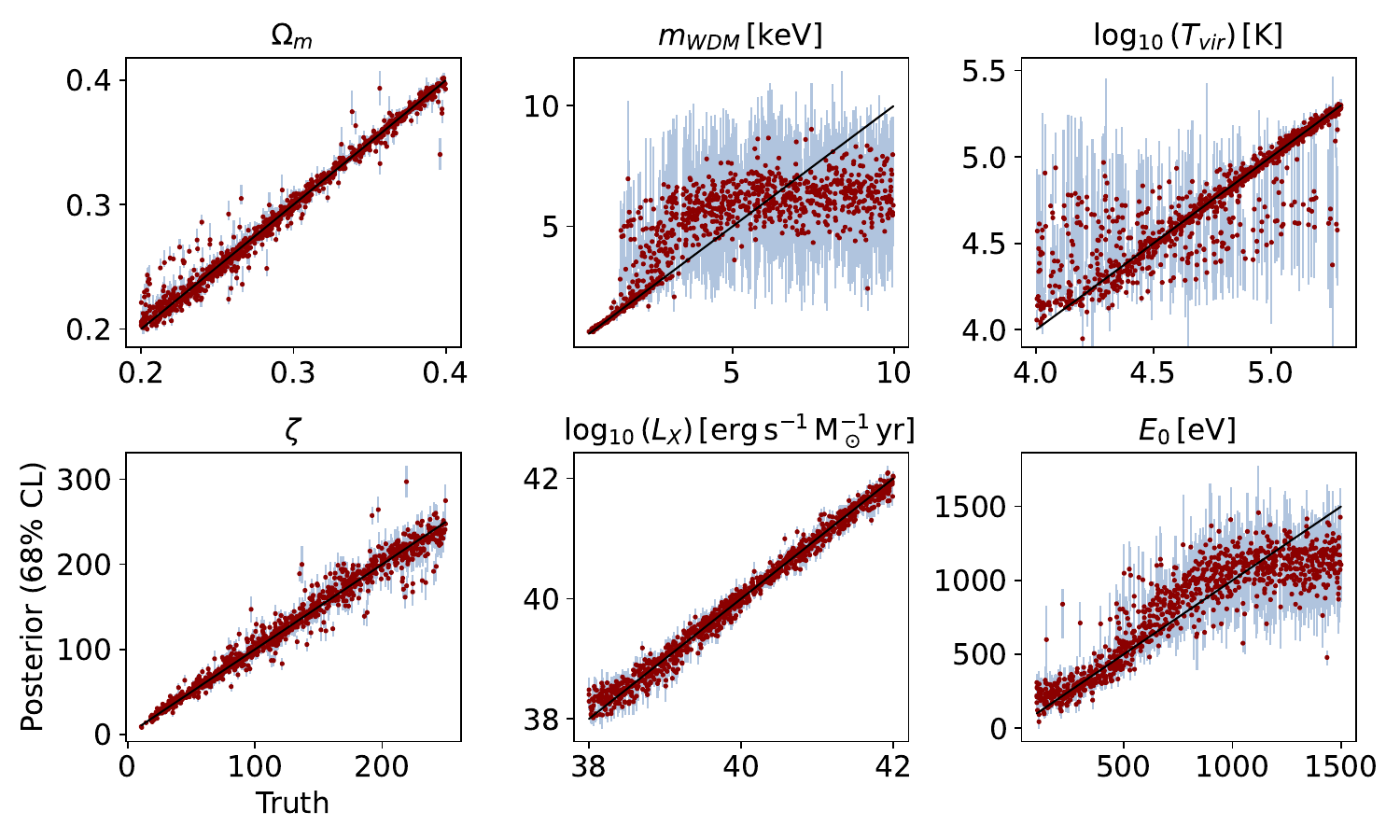}
    \caption{Calibration or parameter recovery for simulations,
      showing the mean of the marginalized posterior (red dots) and
      the 68\% credible intervals (light blue).}
    \label{fig:param_recovery}
\end{figure}

Before looking at the correlated posterior for the combination of
cosmological and astrophysical parameters, we analyze the performance
of the 21cmPIE-INN for the individual marginalized posteriors.  In
Fig.~\ref{fig:param_recovery} we first show the recovered values for our six
key parameters and their marginalized error bars from the cINN for pure simulation. The
errors generally increase away from the diagonal line of perfect
parameter recovery, as expected.  However, for the different 
parameters we see a range of patterns,
from near-perfect recovery to a significant fraction of outliers and
the appearance of degeneracies.

First, the matter density $\Omega_\text{m}$ is extracted almost
perfectly, with small uncertainties and a diagonal calibration
curve. In Tab.~\ref{tab:results_ml} we show the corresponding values
for $R^2$, the NRMSE, and the calibration error of the approximate
posterior.

The most difficult parameter to infer is $m_\text{WDM}$, especially
for large masses. 
First, for small $m_\text{WDM}$ the
network-derived parameter value is perfectly calibrated and very certain. 
For larger masses, the decreasing
free-streaming length of WDM means that the signal looks
more and more like CDM. This means we lose the relevant physics information, so the 
flat calibration curve is not a problem of the summary network or cINN. 
As a matter of fact, the huge error bars on the plateau account for this
loss of information, and a
likelihood analysis using power spectra comes to a similar
result~\cite{saxena}. The metric to check
for the coverage of the posterior, $\text{Err}_\text{cal}(m_\text{WDM})=0.011$ 
is unexpectedly small, because it is not well-suited to capture 
this failure mode, where an overestimation for medium-sized $m_\text{WDM}$ 
and an underestimation for large values cancel each other.

For $T_\text{vir}$ the truth is again recovered well, albeit with a
small group of outliers. They are explained by a degeneracy with
$m_\text{WDM}$~\cite{Neutsch:2022hmv}. A threshold for early star
formation is set by $T_\text{vir}$, but also by
\begin{align}
  M_\text{Jeans} \propto
  (\Omega_\text{m} h^2)^{1/2} \; 
  \left(\frac{m_\text{WDM}}{\text{keV}}\right)^{-4} \; M_\odot \; .
  \label{eq:jeans}
\end{align}
For large $m_\text{WDM}$ the Jeans mass limit becomes more important
and the minimum virial temperature has little effect on the era of
reionization, resulting in a degeneracy for $T_\text{vir}$. However,
the posterior for these parameter combinations is wide enough and does
not underestimate the error budget.

\begin{table}[b!]
\centering
\begin{small} \begin{tabular}{lc|cccccc}
\toprule
&& $\Omega_\text{m}$ & $m_\text{WDM}$ & $T_\text{vir}$  & $\zeta$ & $L_X$ & $E_0$  \\
\midrule
\multirow{3}{*}{Simulation}
& $R^2$                  & 0.981 & 0.621 & 0.764 & 0.969 & 0.987 & 0.803  \\
& NRMSE                  & 0.039 & 0.181 & 0.135 & 0.048 & 0.032 & 0.130  \\
& $\text{Err}_\text{cal}$ & 0.007 & 0.011 & 0.050 & 0.028 & 0.016 & 0.025  \\
\midrule
\multirow{3}{*}{Mock}
& $R^2$                  & 0.990 & 0.663 & 0.716 & 0.969 & 0.973 & 0.045  \\
& NRMSE                  & 0.028 & 0.171 & 0.149 & 0.048 & 0.047 & 0.286 \\
& $\text{Err}_\text{cal}$ & 0.045 & 0.017 & 0.086 & 0.066 & 0.025 & 0.064 \\
\bottomrule
\end{tabular} \end{small}
\caption{Various performance metrics for the 21cmPIE-INN, shown for
  pure simulations and for mock data. $R^2$, NRMSE, and
  $\text{Err}_\text{cal}$ are calculated according to
  Eq.\eqref{eq:R2_determination}, Eq.\eqref{eq:NRMSE}, and
  Eq.\eqref{eq:calibration_error}.}
\label{tab:results_ml}
\end{table} 

Next, $\zeta$ and $L_\text{X}$ 
show almost perfect parameter recovery with small network-derived error bars, as confirmed by 
high $R^2$ values, low NRMSE values, low calibration errors of the approximate
posterior.

Finally, the recovery of $E_0$ degrades towards large values. 
The reason is that $E_0$ describes the threshold of self-absorption for host 
galaxies where this X-ray background is generated by compact X-ray binaries. 
Radiation below this threshold cannot escape the host galaxies. Our prior 
range is deliberately wide 
and motivated by the column density of the interstellar medium (ISM) in simulated high-redshift galaxies~\cite{Das:2017}. 
For large $E_0$, corresponding to high ISM column densities within high-redshift 
galaxies, only a small fraction of X-ray radiation can escape the galaxies, 
leading to a similarly small X-ray heating background for all scenarios above 
around 700\,eV. The fact that high-$E_0$ scenarios become indistinguishable
is reflected in the calibration curve and confirmed by the wide error bars.

One more performance test from Sec.~\ref{sec:inf_valid},
latent space examination is described in the Appendix. 
It checks consistency and convergence,
giving more confidence in the network without having to know the true
parameters. It confirms that the network is well-calibrated and
that we are sampling from the correct posterior.

\subsection{Mock data}
\label{sec:res_mock}

\begin{figure}[t]
  \includegraphics[width=\textwidth]{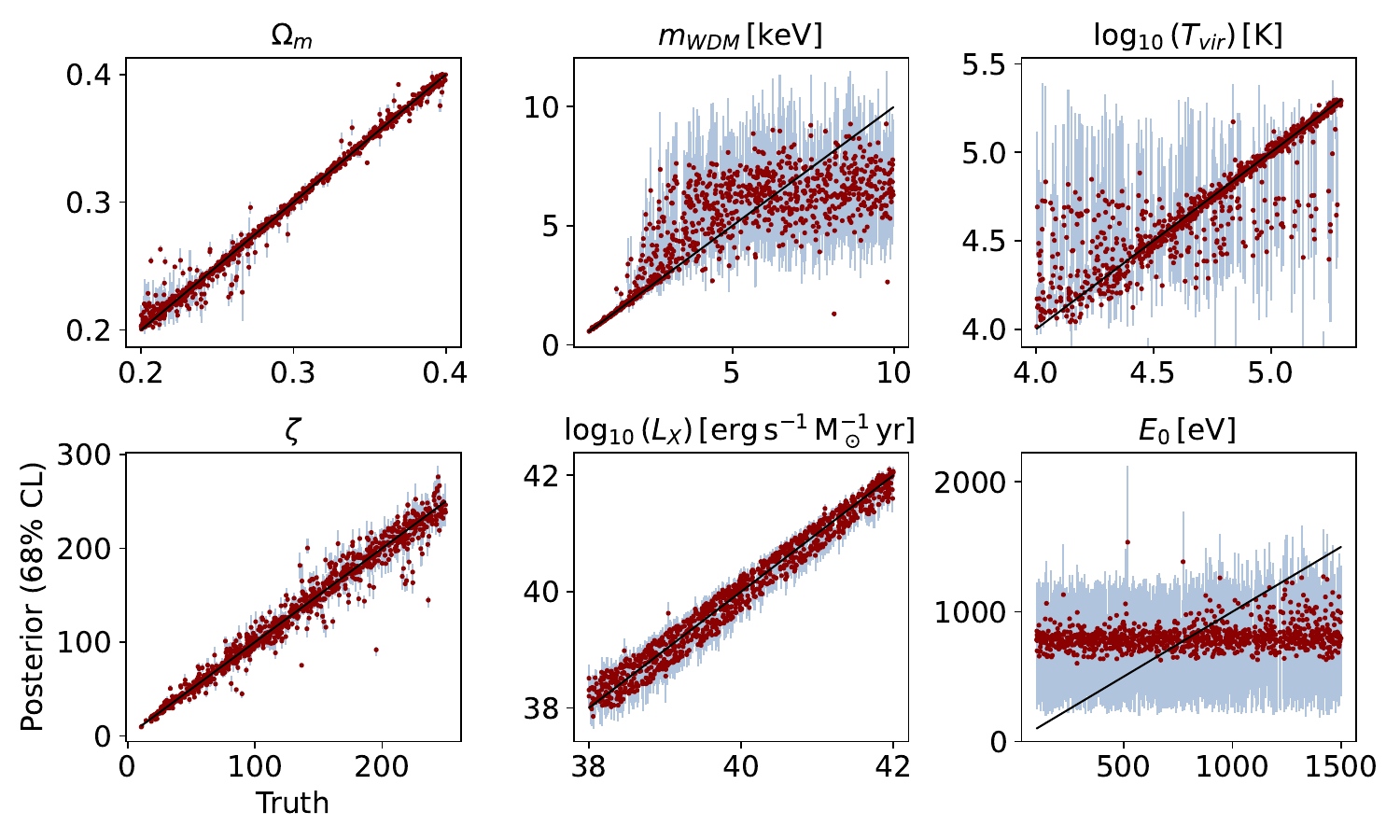}
    \caption{Calibration or parameter recovery for mock data,
      showing the mean of the marginalized posterior (red dots) and
      the 68\% credible intervals (light blue).}
    \label{fig:param_recovery_mock}
\end{figure}
\begin{figure}[b!]
    \includegraphics[width=\textwidth]{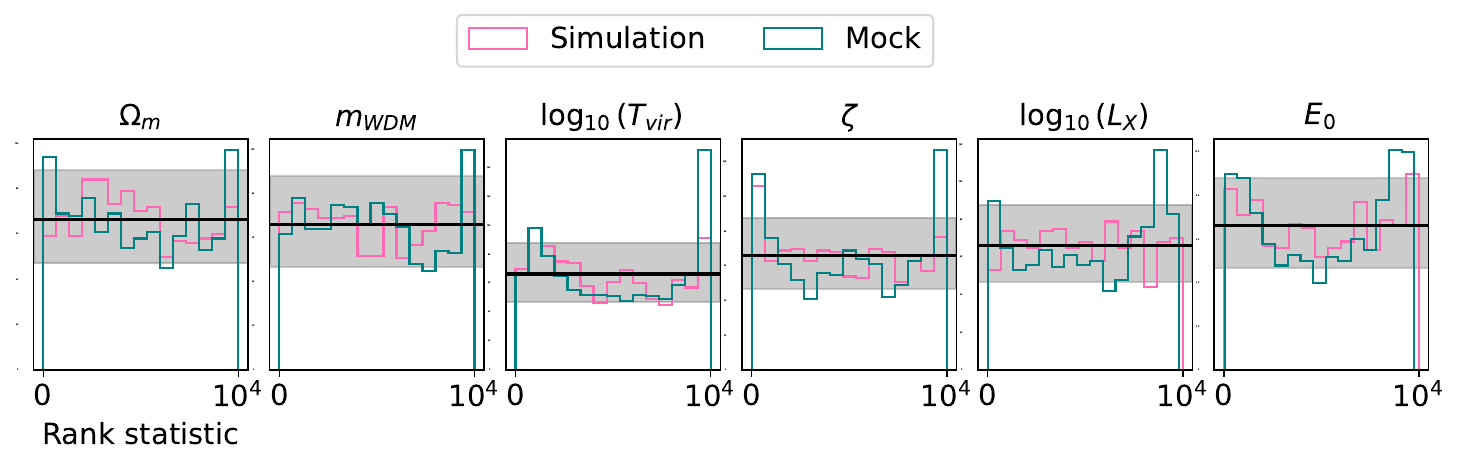}
    \caption{Simulation-based calibration, for pure simulation (pink)
      and mock data (teal).  A uniform distribution indicates no bias,
      with the shaded region indicating the expected variation in
      uniform distributions, here the 99\% quantile of a Binomial.}
\label{fig:sbc}
\end{figure}

It is important to check that our inference does not break down when
the realistic data becomes noisy. This is why we repeat the performance and calibration
study for the mock measurements introduced in Sec.~\ref{sec:inf_data}. These
more realistic results are 
shown in Tab.~\ref{tab:results_ml} and
Fig.~\ref{fig:param_recovery_mock}. The features, challenges, and
patterns are similar to the pure simulation. Especially, the 
inference of the matter density $\Omega_\text{m}$ remains very robust 
and almost perfectly calibrated. This aligns well with previous findings 
that even when transfer learning between different noise levels present 
in 21cm LCs a recovery of $\Omega_\text{m}$ remains feasible~\cite{Neutsch:2022hmv}. 

The main physics effect of the noise added to the mock data is that the network 
now fails to infer $E_0$ altogether. 
This effect is completely unrelated to the effect of noise on the 
network training, discussed in Sec.~\ref{sec:inf_train}. Because it removes 
information from the dataset, it has the opposite effect of the improved 
training performance, and both aspects have to be considered together.
Confidence in the inference method is restored by the fact that the
error bar on the posterior correctly accounts for this. If
anything, the estimated error bars indicate that the network is
slightly under-confident, as can be seen from the low calibration error
$\text{Err}_\text{cal}=0.025$.

The same bottom line can also be extracted from the simulation-based
calibration in Fig.~\ref{fig:sbc}. We show histograms for a sample
size of 10000 and 15 bins. Self-consistent
sampling leads to a uniform distribution. The shaded region is the expected
variation based on the 99\% quantile of a Binomial.  Common failure modes are
overestimation, leading to a $\cup$-shape, and underestimation, leading to a $\cap$-shape.
An asymmetry points towards a systematic bias. In our
case, the histograms for the pure simulation show the expected
variance, implying self-consistent sampling. After adding noise, all
parameters except for $E_0$, and to a much smaller degree $\zeta$, 
remain well-calibrated. For $E_0$ the $\cup$-shape,
implies under-confidence, corresponding to the issues observed in 
Fig.~\ref{fig:param_recovery_mock}. Unlike standard MCMC analysis, where SBC is computationally not feasible, we can claim that our posteriors have the correct coverage and on average no bias.

\subsection{Inference from 21cm light cones}
\label{sec:res_inf}

Finally, we show the full posterior derived for one fiducial model
from the corresponding simulated 21cm LC, with and without
noise. The parameter values of our fiducial
model are
\begin{figure}[t]
    \includegraphics[width=\textwidth]{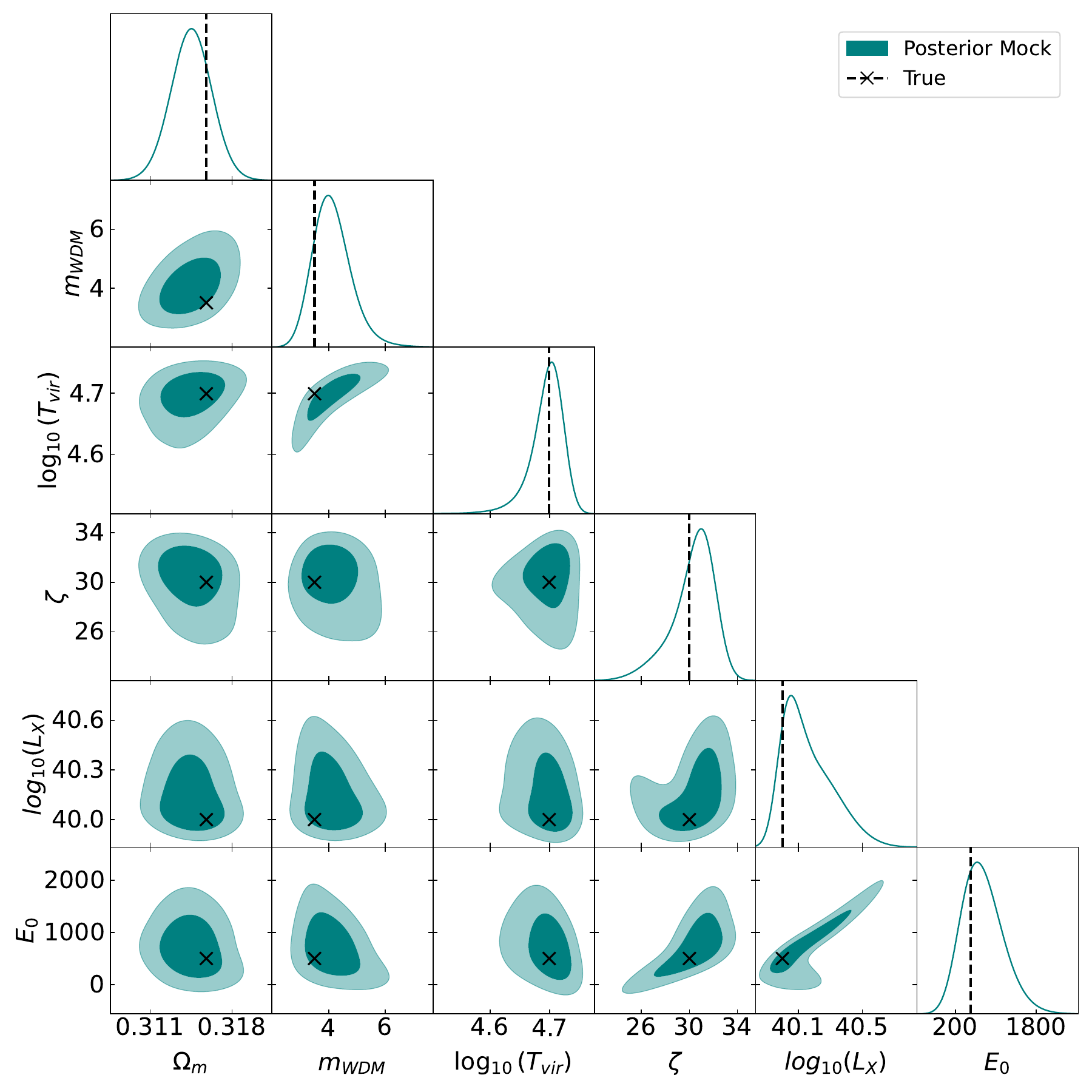}
    \caption{Marginalized posterior from mock measurements and for the light cone given in Eq.\eqref{eq:lc}. 
    The true values are shown as crosses. The shadings indicate 68\% and 95\% CI.}
    \label{fig:post_triangle}
\end{figure}
%
\begin{alignat}{9}
\Omega_\text{m} &= 0.316 
&\qqquad 
m_\text{WDM} &= 3.5 \,\text{keV}
&\qqquad 
T_\text{vir} &= 10^{4.70}\,\text{K} \notag \\
\zeta &= 30
&\qqquad 
L_X &= 10^{40}\,\frac{\text{erg/\text{s}}}{\text{M}_\odot/\text{yr}}
&\qqquad  
E_0 &= 500\,\text{eV} \; .
\label{eq:lc}
\end{alignat} 
Within the prior parameter ranges, any fiducial model parameter set can 
be chosen to extract the posterior fast from the trained model. 
Even though we sample the cINN 100.000 times, to ensure a reliable modeling
of the tails of the multi-dimensional posterior, the analysis of a single 
fiducial takes only a few 
seconds. The numerical bottleneck is loading the data for the light cone.  

The extracted posterior, with 2-dimensional correlations 
and 1-dimensional marginalized curves, is depicted in 
Fig.~\ref{fig:post_triangle}. We only show
the realistic mock data, including noise.
The fiducial parameter values fall within the 1$\sigma$ region for all
parameters.
The posterior for the mock data comes out comparable to what we would get from the 
pure simulation. 
The parameter degeneracies expected from our discussion in Sec.~\ref{sec:res_cal} 
appear in the 2-dimensional correlations. For instance, 
$m_\text{WDM}$ and $T_\text{vir}$ show the degeneracies expected from Eq.\eqref{eq:jeans}. 
Also, the degraded inference of $E_0$ for the mock data can be traced to 
the strong degeneracy with typical X-ray luminosities $L_\text{X}$. 
Most interestingly, the cosmological parameters and especially the matter density 
$\Omega_\text{m}$ are inferred extremely robustly when it comes to adding noise.
More results for different light cones are shown in App.~\ref{app:more}.

A direct comparison with other methods, such as a comprehensive and much slower 
MCMC analysis~\cite{mesinger21cmmc}, is challenging, due to
variations in the included noise models and set of parameters. 
However, our credibility intervals are 
qualitatively similar to those reported in Ref.~\cite{mesinger21cmmc}.

\section{Outlook}

21cm experiments such as the SKA promise precise
measurements of brightness temperature fluctuations of neutral hydrogen, offering a new tomographic perspective
on the high-redshift universe and cosmological structure formation. To make optimal use of this complex and vast dataset, 
we need inference methods beyond simple summary statistics or 
power spectra.
For this purpose, we developed a simulation-based inference method for a joint analysis of 
Cosmology ($\Omega_m$, $m_\text{WDM}$), the Epoch
of Reionization ($T_\text{vir}$, $\zeta$), and Cosmic Dawn ($L_\text{X}$, $E_0$). 
Our modern machine learning setup combines 
an advanced CNN summary network with a
cINN to generate a multi-dimensional correlated posterior by sampling, linked 
by a physics-inspired latent representation of the complex SKA light cones.

We assessed the validity of this inference method through calibration curves 
and a range of metrics,
coefficient of determination ($R^2$) and normalized root
mean square error (NRMSE) for parameter recovery, as well as
simulation-based calibration and calibration error. Aside from 
known correlations, which limit the possible inference, we found exceptional
performance for pure simulation and for mock measurements with added noise.
Only the energy threshold of self-absorption in galaxies $E_0$ becomes a challenge
once noise is added to the dataset, becoming strongly degenerate with the typical 
X-ray luminosity as the second key CD parameter. Notably, the matter density $\Omega_\text{m}$ 
as the key cosmological parameter is robustly inferred even in the presence of noise. 

The summary vector that links the two networks of the 21cmPIE-INN is initialized to the
parameters of interest, but adapted by the joint training with the cINN to guarantee 
an optimal inference, for the given underlying network.  
This optimization, the impact of the form and size of the summary vector, its 
stability, and its benefits in terms of explainable AI is 
an interesting avenue for further investigation. Similarly, our method
allows for training on augmented datasets, for example with different experimental performance, opening further possibilities to analyze realistic SKA data.

Ultimately, the proposed approach facilitates rapid and
straightforward simulation-based inference of likelihood constraints for cosmology. The
generation of full posteriors from a given 21cm light cone takes a
few  seconds, most of this time used for reading in the complex data. 
This speed, combined with the
ability to capture non-Gaussian information and learn optimal summaries (meaning well-calibrated with SBC and on average unbiased), distinguishes this method
from alternative inference approaches.

\subsection*{Acknowledgements}

We would like to thank Theo Heimel for extremely valuable help with INNs and 
Stefan Radev for his support in understanding all things BayesFlow.
BS is funded by Vector Stiftung. CH's work is funded by the Volkswagen Foundation.
This work was supported by the DFG under Germany’s Excellence Strategy EXC
2181/1 - 390900948 \textsl{The Heidelberg STRUCTURES Excellence Cluster}.



\clearpage
\appendix
\section{Latent space examination}
\label{app:latent}

\begin{figure}[hb!]
    \centering
    \includegraphics[width=.8\linewidth]{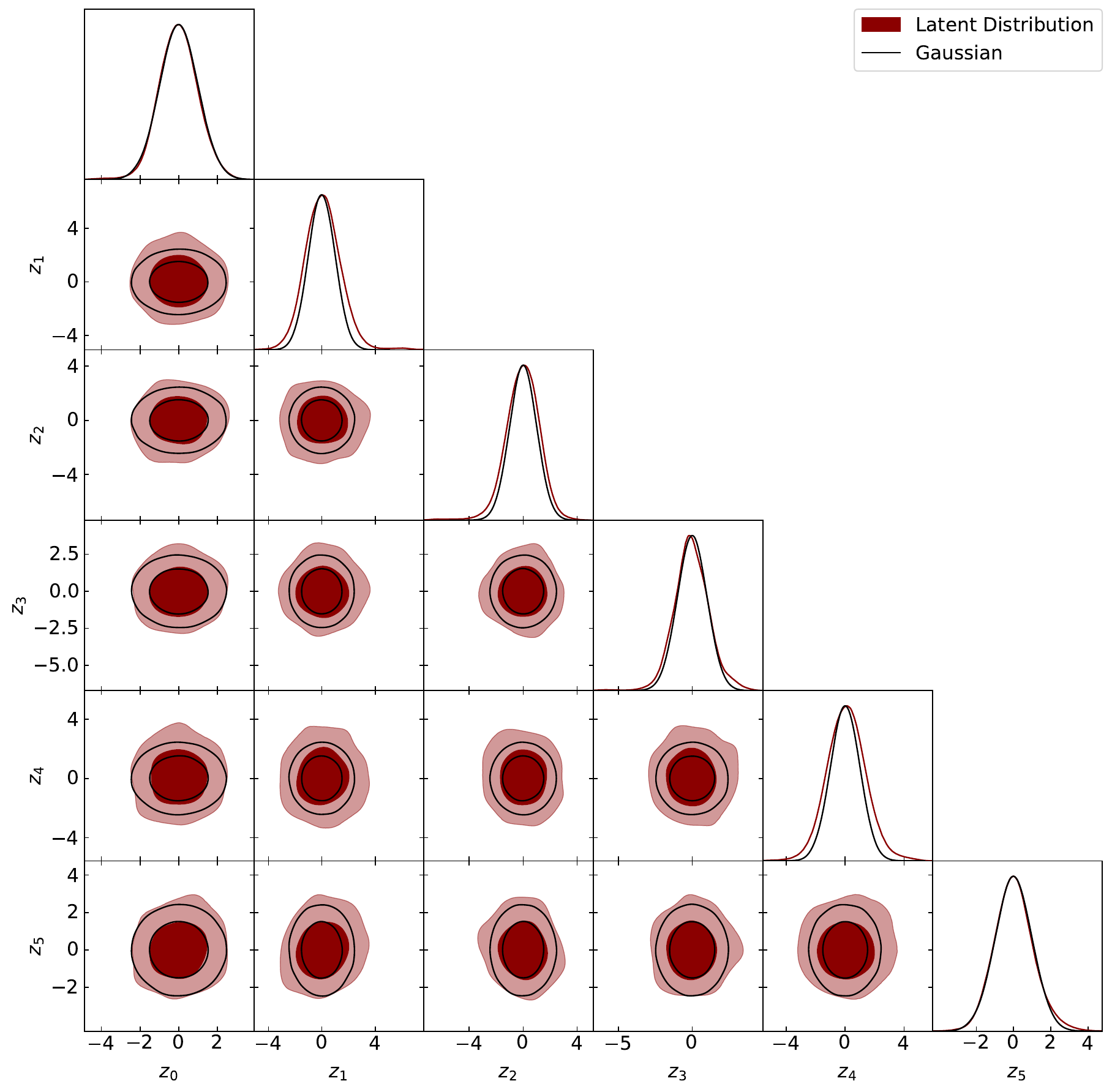}
    \caption{Distribution of the latent space variables and a unit Gaussian as comparison.}
    \label{fig:latent}
\end{figure}

As the learning objective of the cINN is to map the input to normal Gaussian distributions, for a
converged network one would expect it to be distributed as such. This
is quite easy to check visually, by sampling from the latent space and
comparing it to a six-dimensional normal Gaussian. In our two cases of simulations-only and mocks with noise,
both trained cINN networks pass this test. The distribution for the
simulation-only case is shown in Fig.~\ref{fig:latent}.

\section{Further posterior examples}
\label{app:more}

\begin{figure}[hb!]
    \centering
    \includegraphics[width=\linewidth]{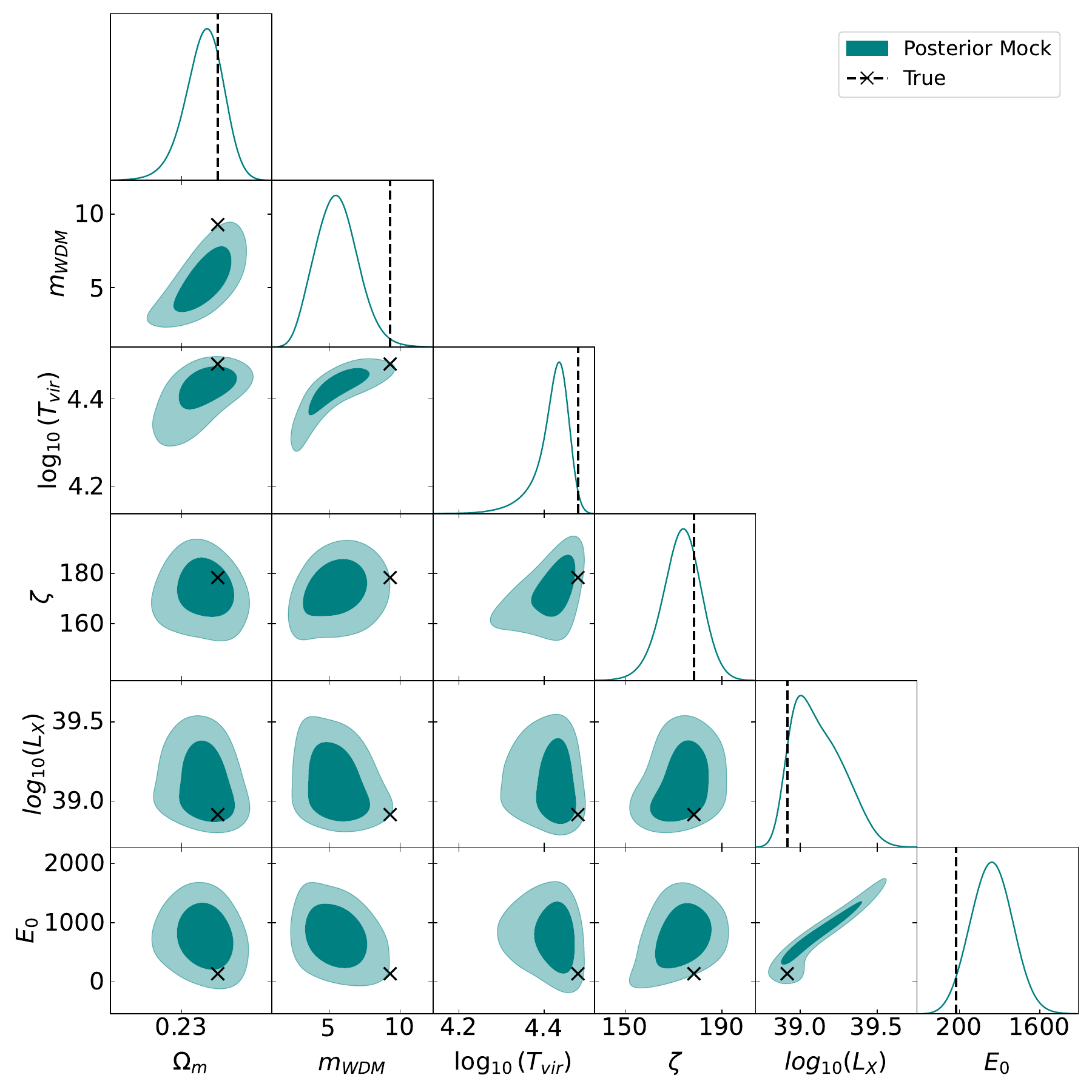}
    \caption{Marginalized 2D posterior and true values ($\Omega_\text{m}=0.237$, $m_\text{WDM}=9.30\,\text{keV}$, $T_\text{vir}=10^{4.48}\,\text{K}$, $\zeta=178$, $L_X=10^{38.9}\,\text{erg}\,\text{s}^{-1}\,\text{M}_\odot^{-1}\,\text{yr}$, $E_0=139\,\text{eV}$) known from the simulation. The contours are from a network trained and evaluated within an optimistic noise scenario for the SKA.}
    \label{fig:post_1}
\end{figure}

\begin{figure}[hb!]
    \centering
    \includegraphics[width=\linewidth]{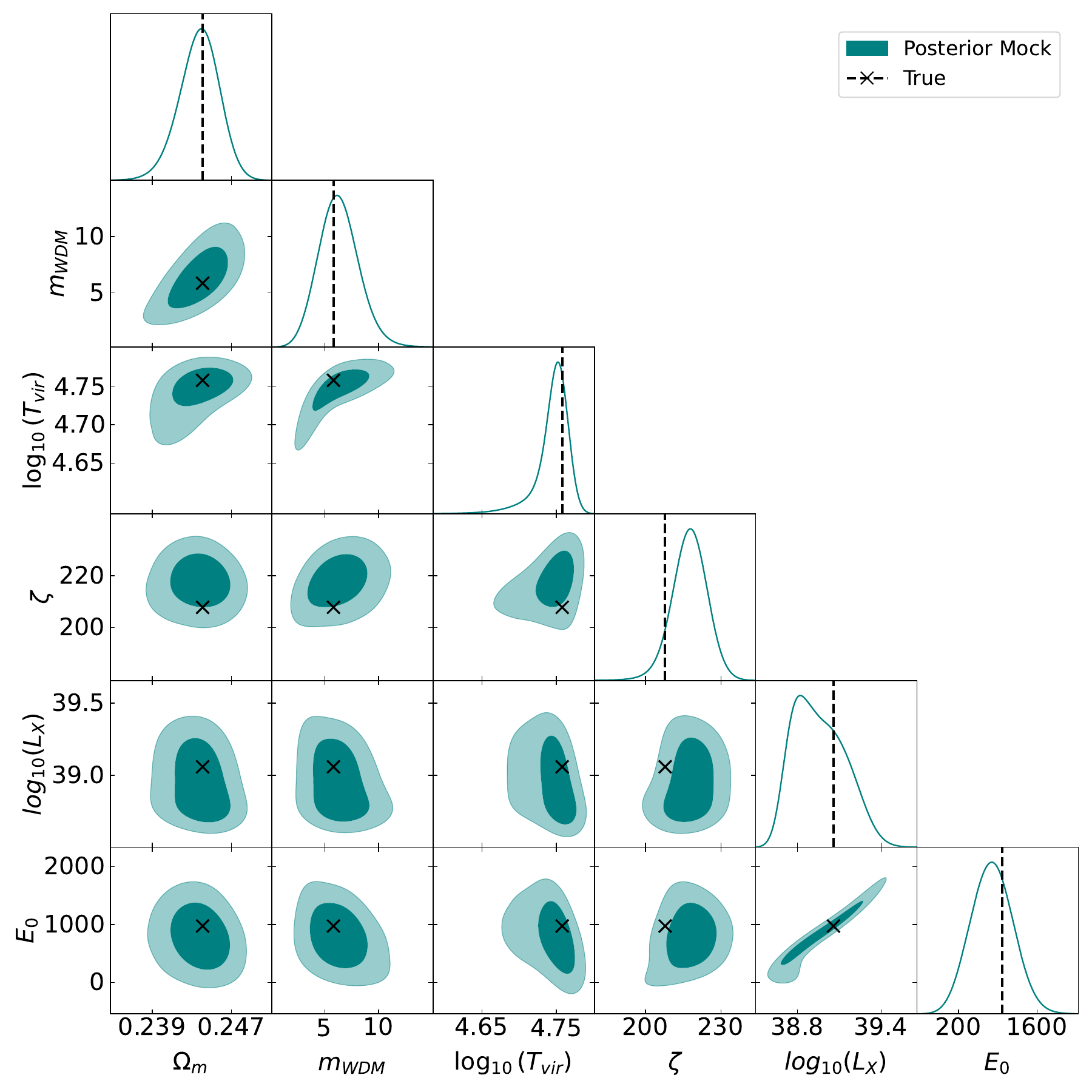}
    \caption{Marginalized 2D posterior and true values ($\Omega_\text{m}=0.244$, $m_\text{WDM}=5.81\,\text{keV}$, $T_\text{vir}=10^{4.78}\,\text{K}$, $\zeta=208$, $L_X=10^{39.1}\,\text{erg}\,\text{s}^{-1}\,\text{M}_\odot^{-1}\,\text{yr}$, $E_0=974\,\text{eV}$) known from the simulation. The contours are from a network trained and evaluated within an optimistic noise scenario for the SKA.}
    \label{fig:post_2}
\end{figure}

\begin{figure}[hb!]
    \centering
    \includegraphics[width=\linewidth]{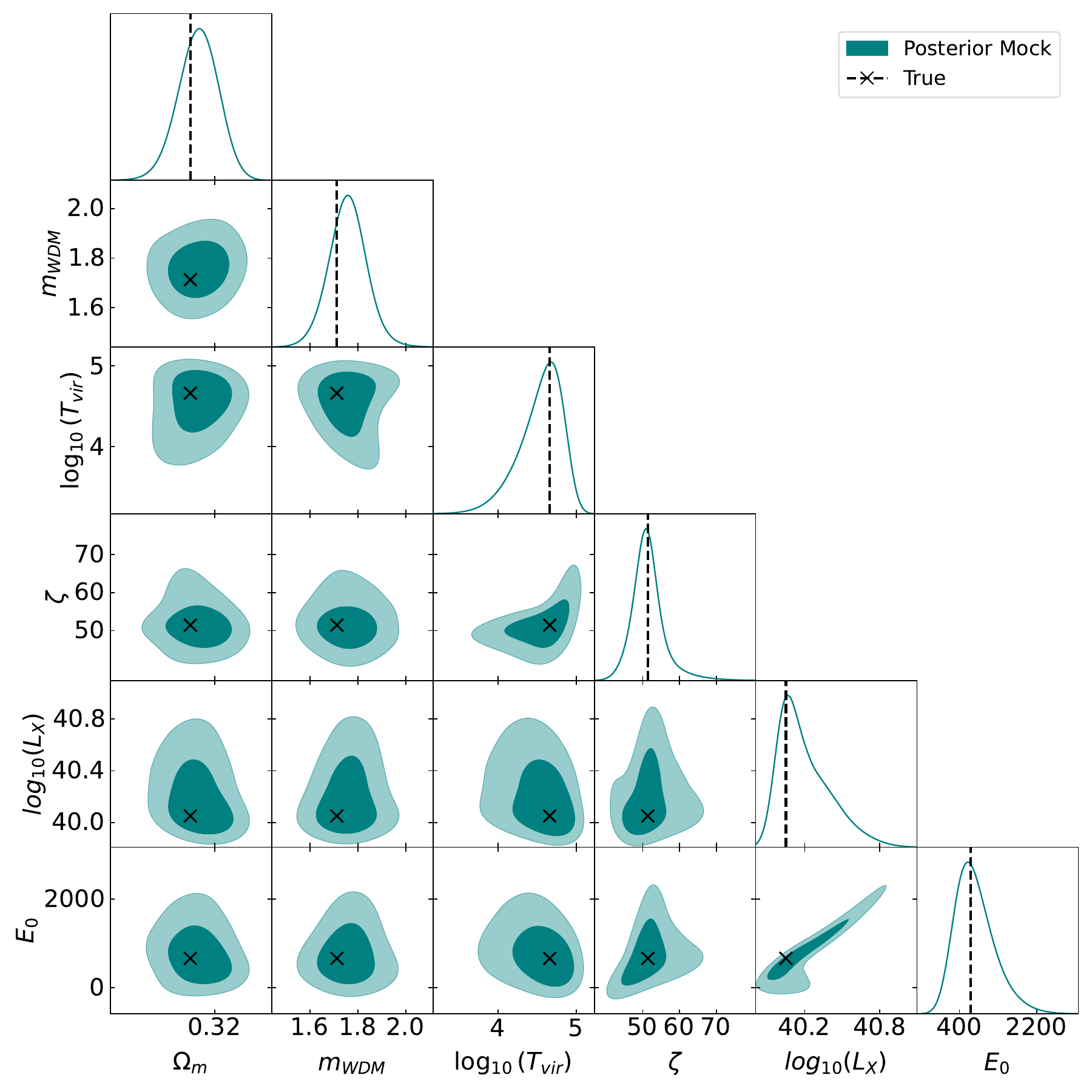}
    \caption{Marginalized 2D posterior and true values ($\Omega_\text{m}=0.318$, $m_\text{WDM}=1.71\,\text{keV}$, $T_\text{vir}=10^{4.66}\,\text{K}$, $\zeta=51.4$, $L_X=10^{40.0}\,\text{erg}\,\text{s}^{-1}\,\text{M}_\odot^{-1}\,\text{yr}$, $E_0=663\,\text{eV}$) known from the simulation. The contours are from a network trained and evaluated within an optimistic noise scenario for the SKA.}
    \label{fig:post_3}
\end{figure}

The fast generation of the full posterior from one fiducial model allows a quick exploration of the parameter space within training prior ranges. 
To illustrate this fast inference from full 21cm LCs, we show three more posteriors for simulation and mock data in 
Figs.~\ref{fig:post_1}-\ref{fig:post_3}. The fiducial parameters are chosen at random and include combinations in the parameter space that are expected to be challenging for inference due to physical reasons, such as \ie large $m_\text{WDM}$. As can be seen from the marginalized 2D posterior contours the inference results are unbiased at the 1-2$\sigma$ level, both with and without noise, and independently of degeneracies or increased errorbars.

\clearpage
\bibliography{references.bib}


\end{document}